\documentclass[prb,aps,twocolumn,showpacs,superscriptaddress]{revtex4-2}
\usepackage{amsfonts,amsmath,bm,multirow}
\usepackage[OT4]{fontenc}

\usepackage{graphicx}
\usepackage{xcolor}
\usepackage{epstopdf}
\usepackage{dsfont}
\usepackage{physics}
\usepackage{hyperref}
\usepackage{lipsum}
\usepackage{dcolumn}

\usepackage{mathtools}
\usepackage{hf-tikz}
\usepackage[normalem]{ulem}

\newcommand{\kp}{k\! \vdot\! p}
\newcommand{\rr}{\bm{r}}

\definecolor{bred}{HTML}{e31a1c}
\definecolor{bgreen}{HTML}{33a02c}
\definecolor{bblue}{HTML}{1f78b4}

\definecolor{armygreen}{rgb}{0.29, 0.33, 0.13}
\definecolor{newred}{RGB}{255,70,70}
\definecolor{newcyan}{RGB}{0,200,255}

\newcommand{\blue}[1]{\textcolor{black}{#1}}
\newcommand{\blueB}[1]{\textcolor{black}{#1}}

\newcolumntype{L}{D{.}{.}{3,4}}

\begin{document}
	
\title {Eight-band $\kp$ description and material gain for selected cubic and pseudo-cubic perovskites}

    \author{Krzysztof Gawarecki}
    \email{Krzysztof.Gawarecki@pwr.edu.pl}
    \affiliation{Institute of Theoretical Physics, Wybrze\.ze Wyspia\'nskiego 27, 50-370 Wroc{\l}aw, Poland}
    \author{Micha{\l} Wi{\'s}niewski}
    \affiliation{Department of Experimental Physics, Wroc\l aw  University of Science and Technology, Wybrze\.ze Wyspia\'nskiego 27, 50-370 Wroc{\l}aw, Poland}
    \author{Maciej Polak}
    \affiliation{Department of Materials Science and Engineering, University of Wisconsin-Madison, Madison, Wisconsin 53706-1595, USA}
    \author{Robert Kudrawiec}
    \affiliation{Department of Semiconductor Materials Engineering, Wroc\l aw  University of Science and Technology, Wybrze\.ze Wyspia\'nskiego 27, 50-370 Wroc{\l}aw, Poland}
    \author{Marta G{\l}adysiewicz}
    \affiliation{Department of Experimental Physics, Wroc\l aw  University of Science and Technology, Wybrze\.ze Wyspia\'nskiego 27, 50-370 Wroc{\l}aw, Poland}
	
	\begin{abstract}
         In this work, we present a ready-to-use symmetry invariant expansion form of the eight-band $\kp$ Hamiltonian for inorganic and organic metal halide perovskites (CsPbX$_3$ and MAPbX$_3$ with $X = \{$Cl, Br, I\}). \blueB{We use the $\kp$ model to calculate} the electronic band structures for perovskite materials of cubic and pseudo-cubic phase. In order to find respective parameters, the band structures of considered materials were \blueB{obtained} within state-of-the-art density functional theory and used next as targets to adjust the $\kp$ bands and determine \blueB{the values of} $\kp$ parameters. The calculated band structure\blueB{s} \blueB{were} used to obtain the material gain for bulk crystals (CsPbCl$_3$, CsPbBr$_3$, CsPbI$_3$, MAPbCl$_3$, MAPbBr$_3$ and MAPbI$_3$) which is compared with the material gain in well-established III-V semiconductors. It was found that for these perovskites a positive material gain appears at lower carrier density than for the reference materials (GaAs and InP). 
         \blue{We demonstrate} that from the point of view of the electronic band structure, the studied perovskites are very promising gain medium for lasers. 
	\end{abstract}
	
	\maketitle
	
	\section{Introduction}
	\label{sec:intr}
    Lasers containing metal halide perovskites as the gain medium have got considerable attention in recent years due to their outstanding emission properties such as high photoluminescence (PL) quantum yield and widely-tuned band gap \cite{Protesescu2015, Nedelcu2015, Dey2018, Lei2021,Han2022,CHEN2023215031}. So far, many papers have been published on two-dimensional (2D or quasi-2D) perovskites where the exciton binding energy is very high and amplified spontaneous emission (ASE) is easily observed in PL measurements \cite{MAUCK2019380,PhysRevMaterials.2.034001,Kahmann2021,Fu2021,WOS:000821277900001,Jin2022,Qin2020}. Significantly less work has been published on lasing and/or ASE in three-dimensional (3D) perovskites \protect\cite{Xing2014,Jia2017,Zhu2015,Pourdavoud2019,Des2014,Geiregat2018,Qin2023}. In this type of perovskites, the exciton binding energy is definitely lower \cite{WOS:000821277900001,Lei2021,Chen2018}. Therefore, \blueB{for some of them} the band-to-band emission can be considered as the main channel of radiative recombination at room temperature, similarly to III-V semiconductors, which are the material base for current semiconductor lasers. This means that the lasing mechanism will be similar to that observed in regular III-V semiconductor lasers, and therefore, it is very interesting to compare these two gain media, i.e., 3D perovskites and III-V semiconductors such as InP or GaAs, in order to assess the prospects of using 3D perovskites in lasers.
  
    Material gain, which quantifies the amount of amplified light per unit length, is a key factor that determines the lasing potential of a given material and can be treated as a figure of merit for designing perovskite lasers. For 3D metal halide perovskites the material gain can be calculated within the $\kp$ method as for regular III-V semiconductors, but such calculations have not been performed so far. \blueB{One should note, that the} gain calculations require accurate modeling of the electronic band structure in the vicinity of the band gap. 
    
    So far, the electronic band structure of perovskite crystals has been extensively studied in the framework of density functional theory (DFT)~\cite{Das2022} as this method is the state-of-the-art approach to studying new materials. \blueB{Such a framework combined with the Bethe–Salpeter equation was also used to obtain absorption spectra for perovskites~\cite{Even2014,Even2015}.} However, the calculation of the material gain by DFT is not developed. Therefore semi-empirical models are used to model the electronic band structure near the band gap with a dense step of the wave vector. The semi-empirical models for calculation of the band structures of bulk perovskites include tight binding ~\cite{Richard2016,Nestoklon2021} and $\kp$~\cite{Yu2016,Aich2020,Fan2018,Ompong2020,Kirstein2022,Sercel2019} approaches. The latter is often used for material gain calculations for ``classical" semiconductors \cite{WOS:000188858600019,WOS:000263161100008,WOS:000294781200020,Gawarecki2022,Scharoch2021}. 
    
\begin{figure}[t]
        \begin{center}
            \includegraphics[width=260pt]{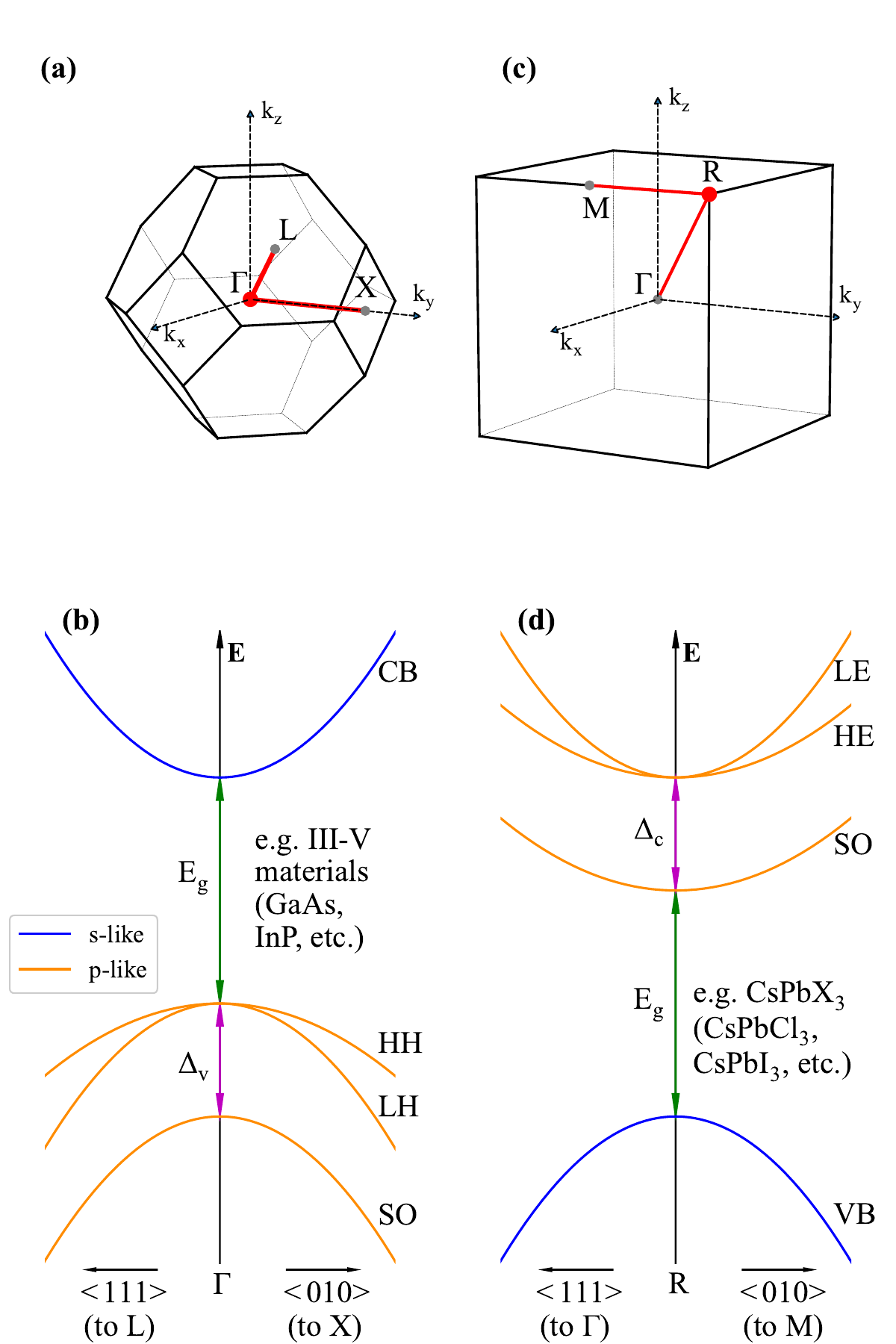}
        \end{center}
        \caption{\label{fig:BZ} Brillouin zone for the zinc-blende (a) and \blue{primitive} cubic (c) crystal lattice, and corresponding to them electronic band structures near the band gap for III-V semiconductors (b) and selected metal halide perovskites (d). One should note, that since directions in (d) are in relation to R-point, paths L-$\Gamma$-X and $\Gamma$-R-M, marked in (a) and (c) are analogous.} 
    \end{figure}

    First of all, it should be emphasized that there are pronounced differences between the electronic band structures for ``classical" gain materials (like InP or GaAs) and perovskites. In the former, the conduction band is composed of the $s$-like states, while the valence band is built on the $p$-like states (with some admixtures allowed by the symmetry)~\cite{LewYanVoon2009}. For the perovskites, the opposite is true, i.e., the valence band is $s$-like and the lowest conduction bands are mostly the $p$-like \cite{PhysRevB.88.165203,AFSARI201710,Even2015} (see Fig.~\ref{fig:BZ}). Furthermore, in contrast to the zinc-blende type structures, for the perovskites the direct band gap often appears at a different point in the Brillouin zone (BZ) than the $\Gamma$ point (\blue{namely} at $R$ for CsPbX$_{3}$)~\cite{Even2015}. For the $\kp$ approach, this requires the formulation of the Hamiltonian for $\bm{k}_0 \neq \bm{0}$ in the BZ (see Fig.~\ref{fig:BZ}). 
    
    To date, several $\kp$ Hamiltonians were proposed for perovskite systems. In Ref.~\cite{Fan2018}, the band structure of cubic-phase CsSnBr$_3$ near the $R$ point of BZ is modeled using the $8$-band $\kp$ model. The same Hamiltonian is used to characterize a wider class of materials (CsBX$_3$ with $B=$\{Pb, Si, Ge, Sn\} and $X=$\{Cl, Br, I\}), and MAPbI$_3$~\cite{Ompong2020}. 
    The symmetry-lowered (C$_\mathrm{4v}$) version of the Hamiltonian was proposed in Ref.~\cite{Yu2016}, where the MAPbI$_3$ in tetragonal phase is modeled. \blueB{The quasicubic $\kp$ Hamiltonian is invoked in Ref.~\cite{Sercel2019}, where the carriers in perovskite nanocrystals are studied.}
    Furthermore, for the tetragonal phase of MAPbI$_3$ and FAPbI$_3$ there is also a 16-band description~\cite{Aich2020}. \blueB{Finally, the $\kp$ model was successfully applied to calculate the Land\`e g-factors in lead halide perovskites~\cite{Kirstein2022}.}
    
    The abovementioned $\kp$ models are expressed either in the JM basis \cite{Fan2018, Ompong2020} or in the (commonly used) basis of S, X, Y, Z states \cite{Yu2016}. However, it would be beneficial to have the \blueB{full} Hamiltonian in an invariant expansion form. The symmetry invariant expansion technique~\cite{Luttinger1956,Trebin1979,LewYanVoon2009} allows writing Hamiltonian in a basis-independent way. This method provides a lot of physical insight, facilitates further calculations (such as perturbation theory), and offers clear transformation rules~\cite{Eissfeller2012}. 
 
    In this paper, we propose an invariant expansion form for the approximated C$_\mathrm{4v}$ Hamiltonian given in Ref.~\cite{Yu2016}.
    Within state-of-the-art DFT calculations, we obtain a set of band structures for inorganic and organic lead halides perovskites of the cubic and pseudo-cubic phase, respectively. We obtain an excellent agreement of the $\kp$ model with DFT results, and thus we determine the $\kp$ parameters for all considered perovskites. Finally, we calculate the material gain for these perovskites, compare it with the material gain for GaAs and InP reference materials, and discuss the prospects of using these perovskites in lasers.

    \section{Material systems}

    \begin{figure}[t]
        \begin{center}
            \includegraphics[width=260pt]{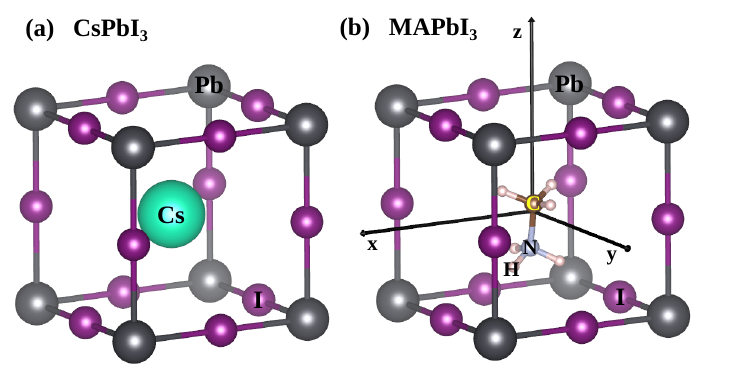}
        \end{center}
        \caption{\label{fig:cell} Atomic arrangement in (a) CsPbI$_3$ and (b) MAPbI$_3$ elementary cell.} 
    \end{figure}
    We consider two classes of perovskite materials: inorganic CsPbX$_3$ and organic MAPbX$_3$ with $X = \{$Cl, Br, I\}. \blue{It is known}, that inorganic and organic halide perovskites can be present in various structures. The phase changes with the temperature, and for many perovskites it is orthorhombic at low temperature \cite{Szafranski2016,Marronnier2018,Kashikar2020}. The transition to the tetragonal phase takes place for increasing temperature, and then to the $\alpha$ cubic (or pseudo-cubic) phase with further temperature increase. For laser applications, it is most interesting to consider room temperature or higher, since semiconductor devices must operate at such conditions. In the case of organic MAPbX$_3$, a pseudocubic phase at room temperature is expected for two of the crystals (MAPbCl$_3$ - above 178.8 K, MAPbBr$_3$ - above 236.9 K, MAPbI$_3$ - above 327.4 K)\cite{1987JChPh..87.6373P,doi:10.1139/v90-063,ONODAYAMAMURO19901383}, while for inorganic CsPbX$_3$, a cubic phase appears for all of them at higher temperatures (CsPbCl$_3$ - above 320.15 K \cite{PhysRevB.9.4549}, CsPbBr$_3$ - above 403.15 K \cite{doi:10.1143/JPSJ.37.1393}, CsPbI$_3$ - above 589.15 K \cite{1361418519902409984, TROTS20082520}).



    Since in this work, we directly compare the results of material gain for different materials, we enforce the $\alpha$-phase for all of them at the same temperature, even though only for two of them (MAPbCl$_3$ and MAPbBr$_3$) pseudo-cubic phase is present at room temperature \cite{Kashikar2020}. 
    
    The CsPbX$_3$ group of materials strictly realizes the cubic symmetry (as shown in Fig.~\ref{fig:cell}(a)) and it is described by the 221 space group (with O$_\mathrm{h}$ point group). 
    
    In the case of MAPbX$_3$, it is known that methylammonium (CH$_3$NH$_3$) fails to match the lattice, causing the symmetry reduction. In consequence, such a group of materials is considered as pseudo-cubic, where approximate cubic symmetry is due to dynamical averaging~\cite{Even2015}. As shown in Fig.~\ref{fig:cell}(b), due to the methylammonium \blueB{(MA)}, the symmetry of the unit cell is completely lifted (the exact symmetry point group of the system is $C_1$). However, if one neglects asymmetry due to the hydrogen atoms and treating N - C pair as localized along the $z$-axis, the cell has $C_{4\mathrm{v}}$ symmetry. Despite the lack of exact cubic symmetry, we describe the MAPbX$_3$ band structures in terms of Brillouin zone for cubic materials [Fig.~\ref{fig:BZ}(c)]. In fact, in Sec.~\ref{sec:kp}, we show that Hamiltonian for the $C_{4\mathrm{v}}$ is capable for accurate description of pseudo-cubic MAPbX$_3$ \blueB{DFT} band structures.

\section{First principles calculations}

The reference band structures were calculated from the first principles within the density functional theory (DFT) \cite{dft:hohenberg_kohn, dft:kohn_sham}. DFT-based first-principles calculations were executed using the Vienna ab initio Simulation Package (VASP) code \cite{dft:vasp1, dft:vasp2}, along with projector augmented-wave (PAW) potentials \cite{dft:paw1}. The selected potentials' valence orbitals were $5$d$^{10}$$6$s$^2$$6$p$^2$, $5$s$^2$$5$p$^6$$6$s$^1$, $3$s$^2$$3$p$^5$, $4$s$^2$$4$p$^5$, $5$s$^2$$5$p$^5$, $2$s$^2$$2$p$^2$, $2$s$^2$$2$p$^3$ for Pb, Cs, Cl, Br, I, C, and N respectively. Because van der Waals interactions have been proven to be critical in accurately defining the geometry and atomic arrangement in halide perovskites \cite{dft:perov_vdw}, the rev-vdW-DF2 functional was utilized for geometry optimization~\cite{dft:vdw-df2}. We optimized lattice vectors and atomic positions using the conjugate gradient algorithm until total energy achieved a convergence of at least of $10^{-5}$ eV/atom, and the maximum residual forces on atoms did not surpass $0.001$ eV/Å. A $6\times 6\times 6$ Monkhorst-Pack \cite{dft:monkhorst_pack} k-point mesh was used in conjunction with a plane-wave basis set defined by a cut-off energy of 600~eV, both values chosen as a result of convergence studies. The electronic band structure was computed using the hybrid HSE06 functional \cite{dft:hse06}. Although hybrid functionals considerably enhance the accuracy of the band gap, it is still slightly underestimated for halide perovskites. To rectify this, the $\alpha$ parameter was adjusted to $\alpha=0.60$ to accurately portray the bandgap. 
Spin-orbit coupling has been included due to its critical role in proper description of the band structure of halide perovskites \cite{dft:spin_perov}. The DFT calculations account for zero-temperature band structures, and the electron-phonon interaction~\cite{Wright2016} is neglected.
\blueB{We provide VASP POSCAR files in the attached Supplementary Material.}

\section{The Hamiltonian in eight-band k.p model}
\label{sec:kp}

\begin{table*}[t]
\centering
\caption{Lattice constants (experimental and calculated) and material parameters for the eight-band $\kp$ model.}
\begin{ruledtabular}
\begin{tabular}{lLLLLLL}
\label{tab:params}
           & \multicolumn{1}{c}{CsPbCl$_{3}$} & \multicolumn{1}{c}{CsPbBr$_{3}$} & \multicolumn{1}{c}{CsPbI$_{3}$} & \multicolumn{1}{c}{MAPbCl$_{3}$} & \multicolumn{1}{c}{MAPbBr$_{3}$} & \multicolumn{1}{c}{MAPbI$_{3}$} \\[0.4ex]
           \hline\\[-1.6ex]
$a_\mathrm{exp}$\hfill(\AA) &  5.605^a &   5.886^b  &6.294^b & 5.675^c & 5.901^c  & 6.329^c   \\
$a_\mathrm{DFT}$\hfill(\AA) &  5.654 &   5.914  &6.293& 5.735& 5.985  & 6.355    \\[0.4ex]
\hline\\[-1.6ex]
$E_{g}$\hfill(eV) &  2.744 &   2.073  &1.416& 3.007& 2.277  & 1.549    \\
$\Delta_{c}$\hfill(eV) & 1.444      &  1.476   & 1.494 &1.506 &1.508  &1.500       \\
$\delta$\hfill(eV) & 0&0    &0  &   0.045 &0.057 & 0.075       \\
$\zeta$\hfill(eV)   & 0 & 0 &0 & 0.016 & 0.030   & 0.045  \\
$P_{||}$  \hfill(eV$\cdot$\AA)  &  9.233  & 8.948  &8.601 &8.878  &8.623        & 8.097   \\
$P_{z}$   \hfill(eV$\cdot$\AA)  &   9.233    &  8.948 &8.601 & 9.896 &9.302        & 9.058       \\
$\gamma_1'$    &   1.643      &  2.183   & 2.997 &1.581 &2.144  &        2.967\\
$\gamma_2'$    &  0.190       &  0.394    & 0.683 &0.140  &0.190      &0.286    \\
$\gamma_3'$    & -0.691        & -1.081    &-1.169  &-0.599 & -1.060       &-1.150 \\[0.4ex]
           \hline\\[-1.6ex]
$m_{v}$         &  0.189     & 0.158   & 0.125 & 0.210 & 0.178  &        0.141\\
$\gamma_1$     &   3.424      & 4.157   & 5.221 & 3.228 &3.954   &        5.001\\
$\gamma_2$     &   1.081      & 1.381   & 1.795  & 0.963   &1.094     & 1.302    \\
$\gamma_3$     &  0.199       & -0.094   & -0.057 & 0.225 &-0.155      &-0.134  \\

\end{tabular}
\end{ruledtabular}
\raggedright$^a$Reference \cite{Moller1958}. \par
\raggedright$^b$Reference \cite{1361418519902409984}. \par
\raggedright$^c$Reference \cite{1987JChPh..87.6373P}. \par
\end{table*}

        In this section, we present an invariant expansion form of the $C_{4\mathrm{v}}$ Hamiltonian for perovskites. This can be used to characterize the materials in tetragonal, pseudocubic, and cubic phase. In the last high-symmetry case, some parameters are equal or set to zero, which makes the Hamiltonian identical (except to the band order) to the well-known form \blue{(see the T$_\mathrm{d}$ Hamiltonian from Refs.~\cite{Winkler2003,Trebin1979,LewYanVoon2009} with the inversion asymmetry terms neglected)}.
	It is convenient to write the Hamiltonian in a block matrix form
	\begin{equation}
		\label{eq:Hblock}
		H=
		\left(
		\begin{array}{*{11}{c}}
			H_{\mathrm{8c8c}} & H_{\mathrm{8c6c}} & H_{\mathrm{8c6v}} \\
			H_{\mathrm{6c8c}} & H_{\mathrm{6c6c}} & H_{\mathrm{6c6v}} \\
			H_{\mathrm{6v8c}} & H_{\mathrm{6v6c}} & H_{\mathrm{6v6v}} 
		\end{array}
		\right),
	\end{equation}
    	where the indices refer to the $\Gamma^-_{8c}$, \blue{$\Gamma^-_{6c}$}, and $\Gamma^+_{6v}$ irreducible representations of the $O_{\mathrm{h}}$ point group. One should note that for the \blue{primitive} cubic system, the R point has the same symmetry as the $\Gamma$ point~\cite{Dresselhaus2010}. However, for the symmetry reduced to $C_{4\mathrm{v}}$, the $\Gamma^-_{8c}$ becomes reducible (which we mark by the symbol $\widetilde{\Gamma}^-_{8c}$) and splits into two irreducible representations following the compatibility relations~\cite{Bir1974,Koster1963}. This results in the lifting of the 4-fold degeneracy in the ``8c'' band block (i.e., splitting the ``heavy'' and ``light'' electrons)~\cite{Yu2016}. Furthermore, the reduction from the $O_{\mathrm{h}}$ to $C_{4\mathrm{v}}$ increases the number of parameters needed to describe the band structure.
	
	We neglect the non-cubic parameters related to far-band contributions~\cite{Yu2016}, keeping only the ones inherently present in the eight-band model. Then, the relevant blocks of the Hamiltonian can be written as 
	\begingroup
	\allowdisplaybreaks
	\begin{align*}
			H_{\mathrm{8c8c}} =&  \qty( E_{\mathrm{g}} + \Delta_{\mathrm{c}} ) \; \mathbb{I}_4  + \frac{\delta}{3} \qty(J^2_z - \frac{J^2}{3}) \\ & + \frac{\hbar^2}{2m_0}  \Bigg \{\gamma'_1 k^2 \; \mathbb{I}_4 - 2 \gamma'_2 \qty[ \qty(J^2_x - \frac{J^2}{3}) k^2_x + \mathrm{c.p.} ] \\& - 4 \gamma'_3 \Big[ \{J_x,J_y\} \{k_x,k_y\} + \mathrm{c.p.} \Big ] \Bigg \} ,\\
			H_{\mathrm{6c6c}} =&  \qty( E_{\mathrm{g}} + \frac{\hbar^2}{2m_0} \gamma'_1 k^2 ) \; \mathbb{I}_2 ,\\
			H_{\mathrm{6v6v}} =&  \frac{\hbar^2}{2m_0} k^2 \; \mathbb{I}_2 ,\\
			H_{\mathrm{8c6c}} =& -\frac{\hbar^2}{2m_0}  \Big \{6 \gamma'_2 \qty[ T^\dagger_{xx} k^2_x + \mathrm{c.p.} ] \\ &+ 12  \gamma'_3 \qty[ T^\dagger_{xy} \{k_x,k_y\} + \mathrm{c.p.} ] \Big \},\\
			H_{\mathrm{8c6v}} =& \sqrt{3} P_\parallel \qty(T^\dagger_x k_x + T^\dagger_y k_y) +   \sqrt{3} \, T^\dagger_z \qty(P_z k_z - i \zeta) ,\\
			H_{\mathrm{6c6v}} =& -\frac{1}{\sqrt{3}}  P_\parallel (\sigma_x k_x+\sigma_y k_y) -\frac{1}{\sqrt{3}} \sigma_z \qty( P_z k_z - i \zeta),			
		\end{align*}
where $\mathbb{I}_n$ are unit matrices ($n \times n$), $J_i$ are $4 \times 4$  matrices of the total angular momentum (for $j=3/2$), $T_i$ are $2 \times 4$ matrices connecting $j=1/2$ and $j=3/2$ blocks, $T_{ij} = T_i J_j + T_j J_i$, and $\sigma_i$ are the Pauli matrices. The remaining off-diagonal blocks are given by the hermitian conjugates, i.e.,  $H_{\mathrm{6v8c}} = H^\dagger_{\mathrm{8c6v}}$, $H_{\mathrm{6v6c}} = H^\dagger_{\mathrm{6c6v}}$, and $H_{\mathrm{6c8c}} = H^\dagger_{\mathrm{8c6c}}$. The explicit form of the matrices can be found in Refs.~\cite{Winkler2003, Gawarecki2022, Trebin1979, LewYanVoon2009}.  The symmetric product is taken as $\{ A,B \} = (A B + B A)/2$. \blueB{The parameter governing the splitting due to the spin-orbit coupling is defined by~\cite{Winkler2003}
\begin{equation*}
\Delta_{\mathrm{c}} = - \frac{3i \hbar}{4 m^2_0 c^2 } \mel{X_c}{\qty[\nabla V \times \bm{p}]_y}{Z_c}.
\end{equation*}}
The parameters related to the interband momentum matrix elements are given by $P_\parallel = \frac{\hbar}{m_0} \mel{S_v}{p_x}{X_c} = \frac{\hbar}{m_0} \mel{S_v}{p_y}{Y_{c}}$ and $P_z = \frac{\hbar}{m_0} \mel{S_v}{p_z}{Z_c}$, while the anisotropy parameters are  
\begin{align*}
\mel{X_c}{H_0}{X_c} &= \mel{Y_c}{H_0}{Y_c} = \frac{\delta}{3},\\
\mel{Z_c}{H_0}{Z_c} &= -\frac{2 \delta}{3}, \\
\mel{S_v}{H_0}{Z_c} &= i \zeta,
\end{align*}
where $H_0  = \frac{\hbar^2}{2m_0}p^2 + V(\rr)$~\cite{Yu2016}. The $\gamma'_{1-3}$ account for the \blueB{influence of} bands that are beyond the $8$-band model.  As mentioned, for such a case, the non-cubic contributions are neglected (see the quasicubic approximation in Ref.~\cite{Yu2016}). 
To obtain a relation to the regular Luttinger parameters, one should include the coupling between the conduction band and valence band blocks. \blue{We neglect here the possible differences between $P_z$ and $P_\parallel$, taking a weighted average $P_\mathrm{av} = (2 P_\parallel + P_z)/3$. This allows us to use the standard perturbative formulas~\cite{Winkler2003}}
\begin{align}\label{f1}
	\gamma_1 &= \gamma'_1 + \frac{2 m_0}{\hbar^2} \frac{P^2_{\mathrm{av}}}{3 \qty(E_{\mathrm{g}} + \Delta_{\mathrm{c}})}, \\ \nonumber
	\gamma_2 &= \gamma'_2 + \frac{2 m_0}{\hbar^2} \frac{P^2_{\mathrm{av}}}{6 \qty(E_{\mathrm{g}} + \Delta_{\mathrm{c}})}, \\ \nonumber
	\gamma_3 &= \gamma'_3 + \frac{2 m_0}{\hbar^2} \frac{P^2_{\mathrm{av}}}{6 \qty(E_{\mathrm{g}} + \Delta_{\mathrm{c}})}. \\ \nonumber
\end{align}
\blue{The electron effective mass in the spin-orbit split-off subband can be related to $\gamma_1$ via~\cite{Vurgaftman2001}}
\begin{align}\label{f2}
	\frac{m_0}{m_\mathrm{so}} &= \gamma_1 + \frac{2 m_0}{\hbar^2}  
	\frac{P^2_{\mathrm{av}} \Delta_{c}}{3 E_{\mathrm{g}} \qty(E_{\mathrm{g}} + \Delta_{c})}. \\ \nonumber
\end{align}

Similarly, one can obtain the valence band effective mass
\begin{align}\label{f2}
	\frac{m_0}{m_\mathrm{v}} &= -1 + \frac{2 m_0}{\hbar^2} \qty( 
	\frac{2 P^2_{\mathrm{av}}}{3 \qty(E_{\mathrm{g}} + \Delta_{c})} + 
	\frac{P^2_{\mathrm{av}}}{3 E_{\mathrm{g}}} ). \\ \nonumber
\end{align}
An explicit form of the Hamiltonian matrix is given in the Appendix~\ref{sec:app_ham}.

\begin{figure*}[t]
    \begin{center}
        \includegraphics[width=.7\textwidth]{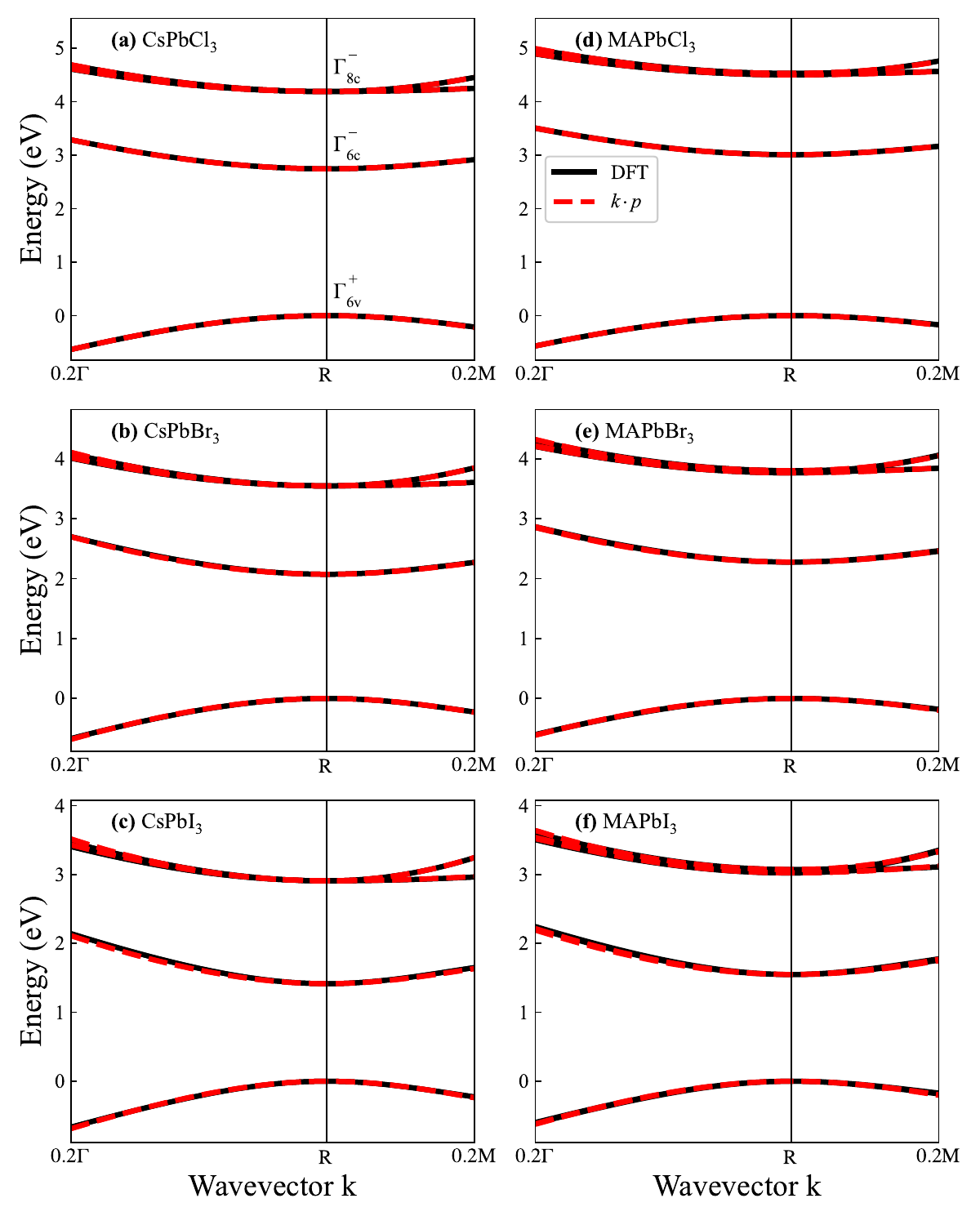}
    \end{center}
    \caption{\label{fig:bands} Comparison of the DFT and the $\kp$ band structures for (a) CsPbCl$_3$, (b) CsPbBr$_{3}$, (c) CsPbI$_{3}$, (d) MAPbCl$_{3}$, (e) MAPbBr$_{3}$, and (f) MAPbI$_{3}$. The ``0.2M" notation means 20$\%$ of the path from point R to point M. } 
\end{figure*}
\begin{figure}[tb]
    \begin{center}
        \includegraphics[width=.48\textwidth]{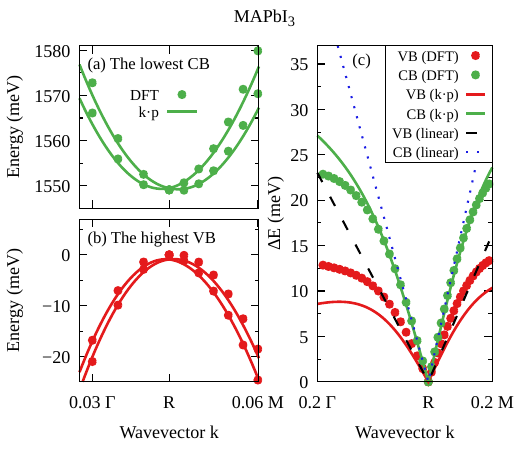}
    \end{center}
    \caption{\label{fig:splittings} \blueB{Comparison of the DFT (points) and the $\kp$ (solid lines) results for MAPbI$_{3}$: (a,b) CB and VB band structures near the R point, (c) the corresponding energy splittings. The dashed lines correspond to the results of the Rashba Hamiltonian (the linear spin splitting). The ``0.2M" notation means 20$\%$ of the path from point R to point M.} }
\end{figure}
We performed fitting for the $\kp$ parameters similarly to the procedure described in Refs.~\cite{Scharoch2021, Gawarecki2022}. The values of E$_{g}$, $\Delta_{c}$, and $\delta$ are extracted directly from the DFT data, i.e., from energies at the $R$ point in the BZ. On the other hand, the values of parameters $\zeta$, $P_{\parallel}$, $P_{z}$, $\gamma_{1}^{'}$, $\gamma_{2}^{'}$, and $\gamma_{3}^{'}$ are obtained from fitting. 
For CsPbX$_3$ the O$_\mathrm{h}$ symmetry point group imposes $\delta = 0$, $\zeta = 0$, and $P_\parallel = P_z$.
All the parameter values are listed in Table~\ref{tab:params}. \blueB{In addition to the $\kp$-related quantities, we also show the values of the lattice constants calculated from the DFT ($a_\mathrm{DFT}$) and their experimental values ($a_\mathrm{exp}$).} For reference purposes, we also list
$\gamma_{1}$, $\gamma_{2}$, $\gamma_{3}$, and m$_{v}$, which are calculated from Eqs.~\ref{f1} and Eq.~\ref{f2}. \blue{In the Appendix~\ref{sec:app_mass}, we compare $m_\mathrm{v}$ calculated in this way to the values obtained from a direct fitting to the DFT results.}

We calculated band structures for all considered materials using the presented $\kp$ model with the obtained parameter sets. \blue{As the deviation from the cubic symmetry is relatively small, we characterize the band structures for MAPbX$_3$ using the notation from the cubic BZ.}
As shown in Fig.~\ref{fig:bands}, we get an excellent agreement between the DFT and the $\kp$ models in a considerably wide range of the BZ. In the case of MAPbX$_3$, the 4-fold degenerated at the R point (for the exact cubic symmetry) block $\widetilde{\Gamma}^-_{8c}$ splits into two blocks of energies $E_{\mathrm{hc}}$ and $E_{\mathrm{lc}}$~\cite{Yu2016}. They are double degenerated due to spin (the Kramers' degeneracy). The splitting between them is given by $E_{\mathrm{hc}} - E_{\mathrm{lc}} = 2 \delta/3$. Since it is not very large (a few dozen of meV), the splitting is not visible in the figure.

\blueB{To check the accuracy of the $\kp$ model for calculating more subtle effects caused by the symmetry lowering, we examine the spin splittings in the highest valence band and the lowest conduction band. Among the considered band structures, for MAPbI$_{3}$ we have the most pronounced deviation from the ideal cubic symmetry (the largest $\zeta$ and $\delta$ parameters), therefore it has been chosen for the model comparison. The results are given in Fig.~\ref{fig:splittings}. As one can see, the agreement is very satisfactory, so the eight-band $\kp$ is capable of handling this effect accurately. The energy splittings for the VB ($\Gamma^+_{6v}$) and the CB ($\Gamma^-_{6c}$)  are also compared with the results of the $2$-band Rashba Hamiltonian giving (see Appendix~\ref{sec:app_rashba}) $\Delta E_\mathrm{c/v} = 2 \abs{\alpha_\mathrm{c/v}} \sqrt{k^2_x + k^2_y}$, where 
\begin{align*}
\alpha_\mathrm{v} &= -\frac{2 \zeta P_\parallel \Delta_\mathrm{c} }{3 E_\mathrm{g}(E_\mathrm{g} + \Delta_\mathrm{c})} \\
\alpha_\mathrm{c} &= -\frac{2 \zeta P_\parallel}{3 E_\mathrm{g}}.
\end{align*}
The formulas for $\alpha_\mathrm{v/c}$ correspond to these presented in Ref.~\cite{Yu2016} for the limiting case of $\delta \ll \Delta_\mathrm{c}$.
One should note that the Rashba parameter $\alpha_\mathrm{v}$ depends on the spin-orbit coupling and vanishes for $\Delta_\mathrm{c} = 0$. In contrast, the linear splitting in the $\Gamma^-_{6c}$ is caused only by the symmetry breaking (here represented by $\zeta$)~\cite{Yu2016}. This is due to different symmetries of the involved Bloch states.
As one can see in Fig.~\ref{fig:splittings}(c), the linear model gives a good agreement with the results of the eight-band $\kp$ in the vicinity of the R point. The values calculated for MAPbI$_3$ are $\abs{\alpha_\mathrm{v}} = 0.0823$~eV\AA \, and $\abs{\alpha_\mathrm{c}} = 0.162$~eV\AA, which is close to the reported DFT values ($0.1 - 0.2$~eV\AA \, range for various MA orientations in the unit cell)~\cite{Frohna2018}. Since our work is devoted mainly to material gain, we limited the calculations to a single MA orientation [as shown in Fig.~\ref{fig:cell}(b)].}
%

\section{ Material gain calculation}
	\begin{figure}
		\begin{center}
			\includegraphics[width=.48\textwidth]{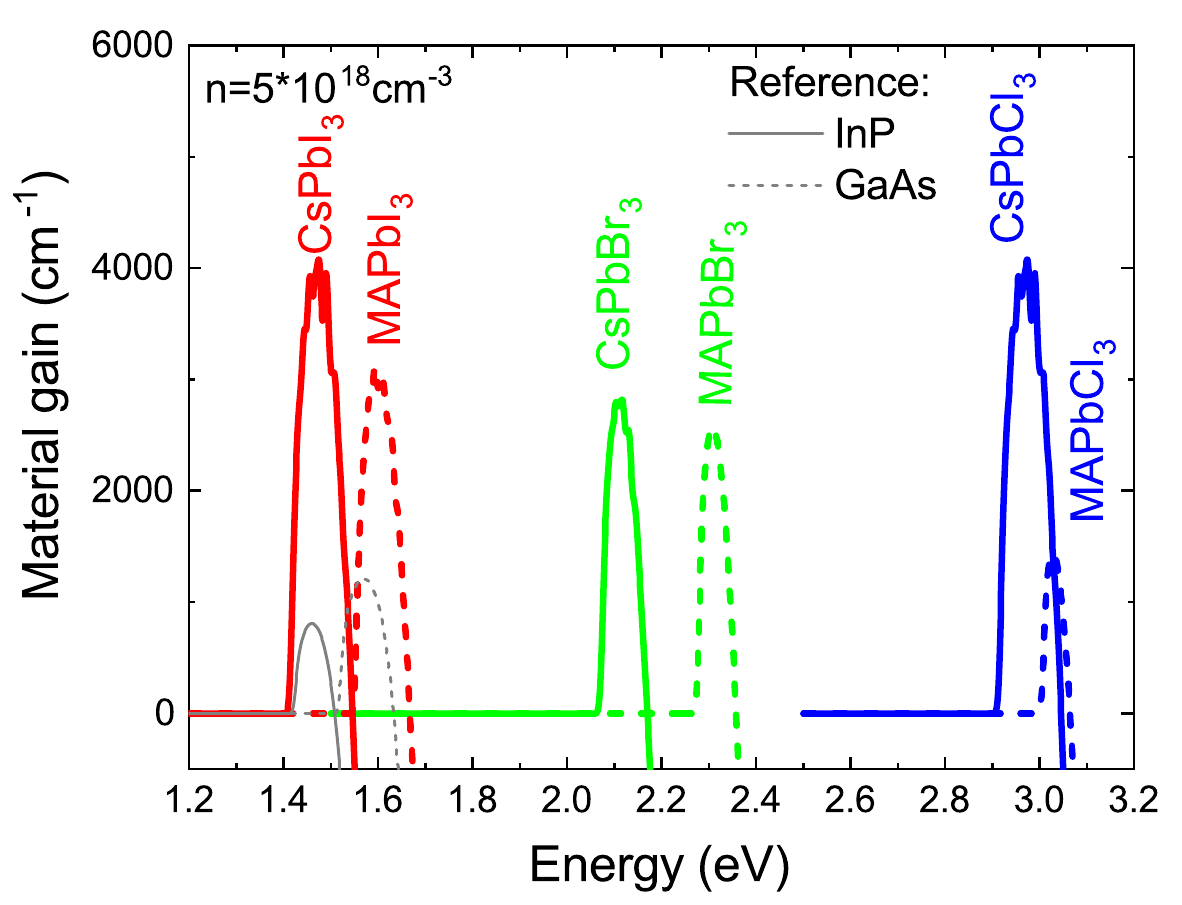}
		\end{center}
		\caption{\label{fi3} Material gain calculated for CsPbCl$_{3}$, CsPbBr$_{3}$, CsPbI$_{3}$, MAPbCl$_{3}$, MAPbBr$_{3}$, and MAPbI$_{3}$ within the eight-band $k\cdot p$ model and taking carrier concentrations 
  \blueB{ $5.0\cdot 10^{18}\mathrm{cm}^{-3}$}. The reference materials are InP and GaAs.} 
	\end{figure}
	\begin{figure*}[t]
		\begin{center}
			\includegraphics[width=.75\textwidth]{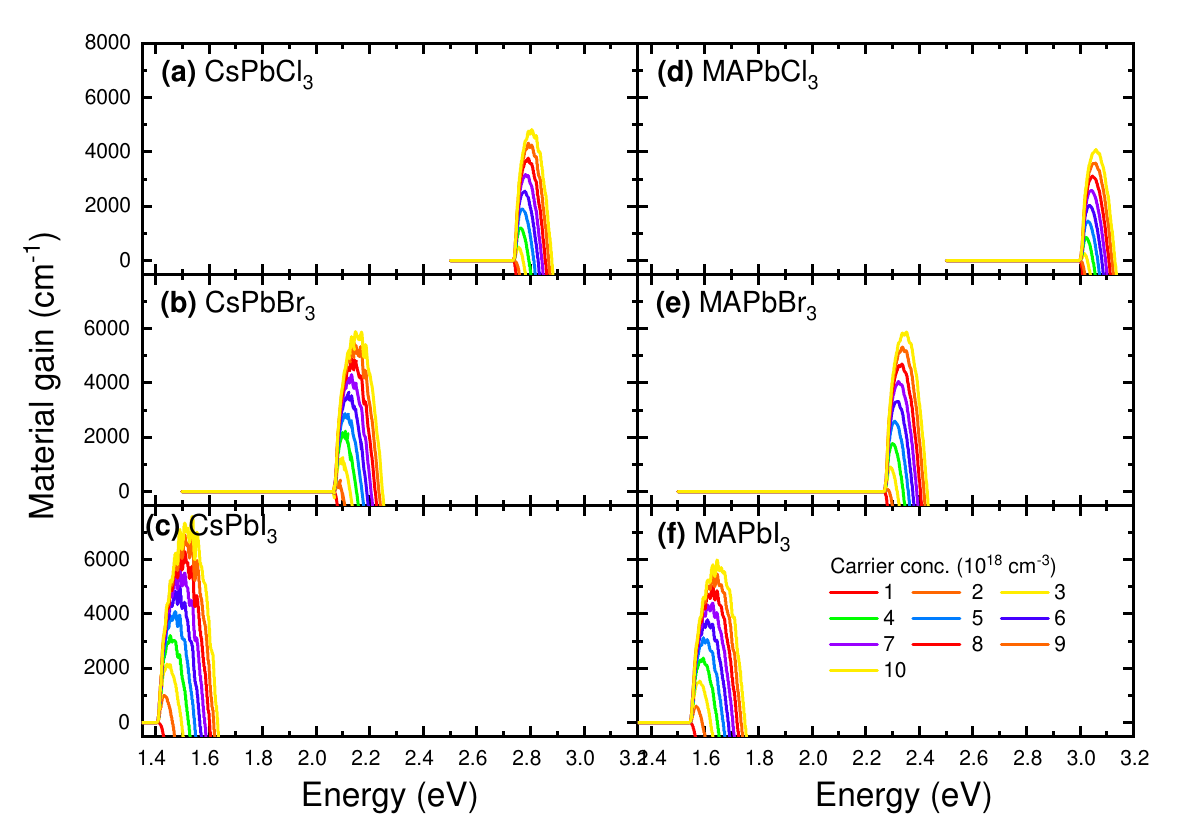}
		\end{center}
		\caption{\label{fig2} Material gain calculated for (a) CsPbCl$_3$, (b) CsPbBr$_{3}$, (c) CsPbI$_{3}$, (d) MAPbCl$_{3}$, (e) MAPbBr$_{3}$, and (f) MAPbI$_{3}$ within the eight-band $k\cdot p$ model and taking carrier concentrations from \blue{$1.0\cdot 10^{18}\mathrm{cm}^{-3}$ to \blueB{$10\cdot 10^{18}\mathrm{cm}^{-3}$}}. The same colors correspond to a given carrier concentration. } 
	\end{figure*}
	\begin{figure}
	   \begin{center}
		\includegraphics[width=.48\textwidth]{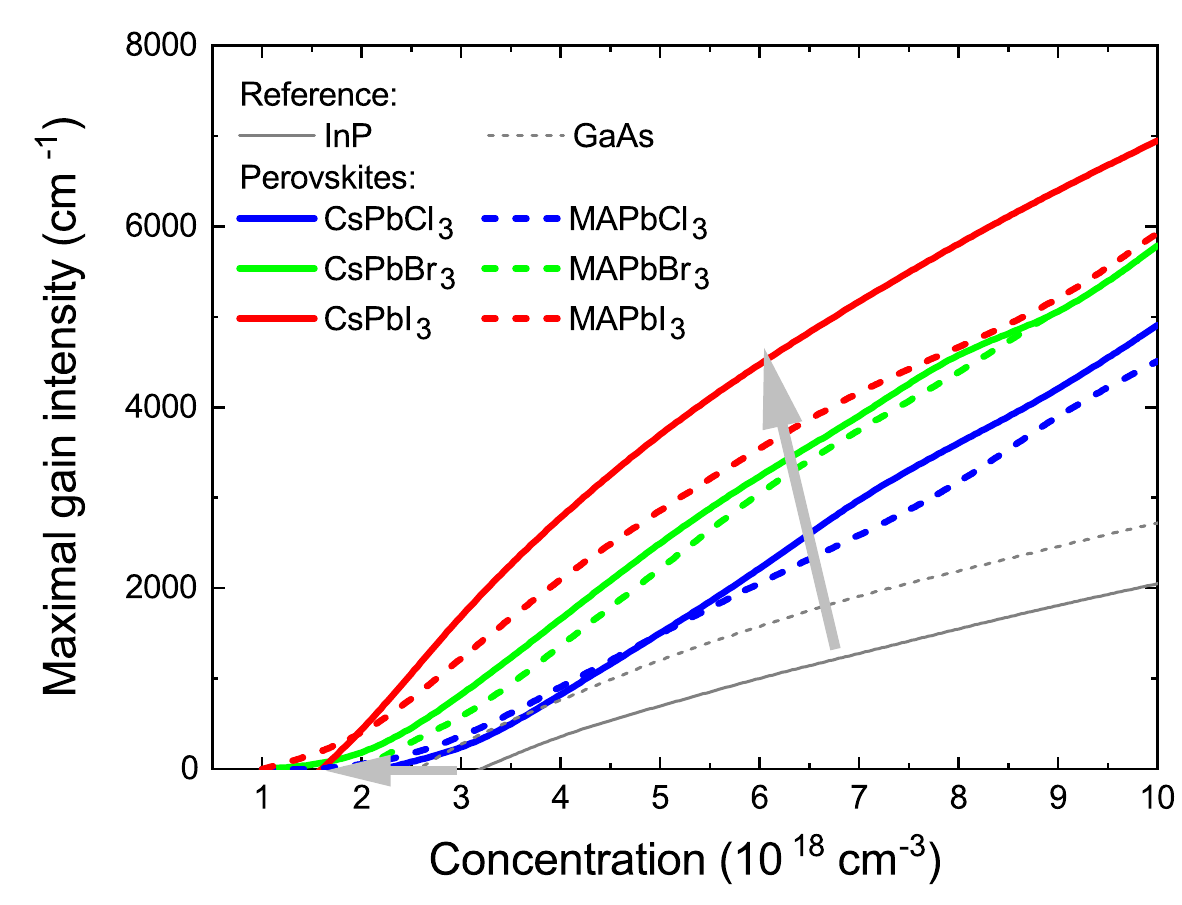}
		\end{center}
		\caption{\label{fig4} Maximal gain intensity dependence on carrier concentration for inorganic (thick solid lines), organic (thick dashed lines) \blueB{perovskites}, and reference III-V materials (thin solid and dashed lines). } 
	\end{figure}
We calculate the material gain for bulk perovskites and compare the results to the values from reference III-V materials: InP and GaAs. 
Details regarding the model for gain calculation are given in Refs.~\cite{
Ell:89,Scharoch2021,Gawarecki2022}. 
The gain was modeled at room temperature according to the equation
\begin{align*}
    \label{Eq:gain}
     g(E)&=\left ( \frac{\Delta k }{2\pi} \right )^{3}
    \frac{\pi e^{2}}{c \epsilon_{0}m_{0}^{2}n_{r}}
\sum_{\bm{k}}\sum_{c,v}
     |\langle c\bm{k}|\hat{\bm{e}}\hat{\bm{p}}|v\bm{k}\rangle|^2 \times\\
    &\frac{\hbar}{\Delta E_{cv;\bm{k}}}\;
\left [ f(E_{c;\bm{k}},F_{c})- f(E_{v;\bm{k}},F_{v})\right ] G(\Delta E_{cv;\bm{k}},E, \gamma),
\end{align*}
 where  $\Delta k$ is a step in the $\bm{k}$-space,  $\epsilon_0$ is the vacuum permittivity, $m_{0}$ is the free electron mass and $n_{\mathrm{r}}$ is the refractive index. The summation indices $c$ and $v$ denote subbands within the conduction and valence band with the respective energies $E_{c;\bm{k}}$ and $E_{v;\bm{k}}$. The energy difference at a given $\bm{k}$ point is $\Delta E_{cv;\bm{k}} =  E_{c;\bm{k}} - E_{v;\bm{k}}$. The f(E$_{c;\bm{k}}$, F$_{c}$) and f(E$_{v;\bm{k}}$, F$_{v}$) are Fermi distribution functions for electrons and holes, respectively. They are calculated for given carrier density and give us quasi-Fermi levels F$_{c}$ and  F$_{v}$.
 To model the allowed transitions, we take the Gaussian distribution function
 \begin{equation*}
   G(E_0, E, \gamma  )=\frac{\sqrt{8\ln{2} }} {\gamma \sqrt{2\pi}} \exp{-\frac{(E-E_0)^{2}8\ln{2}}{2\gamma^{2}}}
  \end{equation*}   
 with a full width at half maximum \blueB{$\gamma=0.01$~eV}. The Gaussian distribution function is usually used for non-Markovian gain calculations \cite{Ahn1995,Park2007}. This kind of line shape gain model gives a good
agreement with experimental data for inhomogeneous systems including alloys and quantum wells \cite{Park2000,Glad2023} and seems to be the most appropriate for perovskites at room temperature as the broadening of PL is also Gaussian like at room temperature \cite{Wehrenfennig2014}. For reference III-V materials, a Lorentzian broadening was used in previous studies \cite{Scharoch2021,Gawarecki2022}. In this case, we calculated the material gain using both Lorentzian and Gaussian broadening functions, but for direct comparison with perovskites, we used Gaussian broadening.
\section{Results and discussion}
	\label{sec:results}
Figure \ref{fi3} shows a direct comparison of material gain spectra calculated for the same carrier concentration of \blueB{$5\cdot 10^{18}\mathrm{cm}^{-3}$} for inorganic (thick solid lines) and organic (thick dashed lines) halide perovskites as well as the InP (thin solid line) and GaAs (thin dashed line) reference materials. \blue{The value of concentration in such a range of values is typically taken into account for the calculation of gain in structures built on the basis of GaAs \cite{GaAsc1,GaAsc2,GaAsc3} and InP \cite{InPc1,InPc2,InPc3}. In the case of CsSnI$_3$ perovskite, the carrier concentration at room temperature was estimated as $10^{17} \mathrm{cm}^{-3}$ \cite{Chung2012,Huang2014}. However, one can expect that for pumped systems relevant for the lasing actions, the concentration gets larger~\cite{Qin2021}. The comparison in Fig.~\ref{fi3} shows} that for inorganic and organic perovskites it is possible to obtain material gain comparable to that observed for typical laser materials such as InP or GaAs. It is also worth noting that in the case of reference materials, very similar gain spectra were obtained for Lorentzian broadening (not shown in Fig. \ref{fi3}), which can be more appropriate in this case. It is worth noting that the electron-phonon coupling in III-V materials is weaker than in metal halide perovskites. The latter are considered to be very soft materials~\cite{Munson2018, WOS:000738134000001}, and thus are more thermally inhomogeneous at room temperature than III-V materials, and therefore, Gaussian broadening is selected for this comparison.

\blue{We performed further}  calculations of the material gain as a function of carrier concentration and the intensity of the gain peak was carefully analyzed. Figure~\ref{fig2} shows the results of material gain calculations for inorganic (left panel: CsPbCl$_{3}$, CsPbBr$_{3}$ and CsPbI$_{3}$) and organic (right panel: MAPbCl$_{3}$, MAPbBr$_{3}$ and MAPbI$_{3}$) perovskites. They are obtained for carrier concentrations from \blue{$1\cdot 10^{18}\mathrm{cm}^{-3}$ to \blueB{$10\cdot 10^{18}\mathrm{cm}^{-3}$}}. For all perovskite crystals, the material gain spectrum changes very similarly: an increase in material gain and a blue shift of gain peak is observed with the increase in carrier concentration. The maximum values of the gain as a function of carrier concentration are shown in Fig. \ref{fig4}. In this case, the gain peak intensity is also plotted by thin lines for reference materials. \blue{One can} conclude that \blue{from the point of view of} the electronic band structure, the considered perovskites \blue{are} very favorable for laser applications because the positive material gain occurs at a lower carrier density than in InP or GaAs, and the gain value is stronger with the same carrier concentration, see the gray arrows in Fig. \ref{fig4} for looking trends. 

Also, quite significant differences are observed within the two sets of perovskites, and a clear chemical trend can be found in this case. The highest gain values are obtained for X = I materials. As shown in Table~\ref{tab:params}, CsPbI$_{3}$ and MAPbI$_{3}$ have the smallest effective masses for the valence band (m$_{v}$), i.e., 0.125 and 0.141, respectively. The intermediate values of material gain are obtained for X = Br materials, which have the masses of 0.158 for CsPbBr$_{3}$ and 0.178 for MAPbBr$_{3}$. The highest m$_{v}$ values of 0.189 and 0.210 are for CsPbCl$_{3}$ and MAPbCl$_{3}$, respectively, which correlate with the lowest gain value. 

Differences in \blue{the curvature of the bands  (i.e., the effective masses) }
are one of the reasons for the differences in the gain spectra.  Another reason is related to the band types involved in the process.
For the reference materials, the fundamental transition is between the s-like conduction band and the p-like heavy-hole/light-hole valence band, while for the considered perovskites the fundamental transition is between the p-like spin-orbit split-off conduction band and the s-like valence band, as shown in Fig.~\ref{fig:BZ}. \blue{One should note}
that higher gain values for perovskites are also associated with a lower value of the refractive index compared to the refractive index of the reference materials. \blue{The values of refractive index taken in our calculations are shown in Table~\ref{tab:eps}.}
\begin{table}
    \label{tab:eps}
    \centering
    \begin{tabular}{|c|c |}
    \hline
    Material & $n_{r}$ \\
    \hline
        CsPbCl$_3$ & 2.30 \\
        CsPbBr$_3$ & 2.30\\
        CsPbI$_3$  &2.20\\
        MAPCl$_3$ &2.75\\
        MAPBr$_3$ &2.15\\
        MAPbI$_3$ &2.75 \\
        \hline
        GaAs &3.95 \\
        InP &3.59 \\
        \hline
    \end{tabular}
    \caption{ \blue{The values of refractive index from Ref.~\cite{refIdx}}}
 \end{table}

The quantitative differences in gain measurements of perovskites and III-V materials may be influenced by several other factors, including \blue{the difference in carrier concentrations. Furthermore, the outcome can be affected by } quality of the material, which can be strongly manifested in experimental results, and which is not included in our calculations. Nevertheless, the calculations and analysis carried out in this article allow us to unequivocally conclude that from the point of view of the electronic band structure, the studied perovskites are excellent gain medium for lasers. From the point of view of material quality and stability, there are many challenges that need to be overcome, but they are not addressed in this article. 

At present, direct comparison of gain measurements with our theoretical predictions are limited as most experimental studies focus on 2D perovskites, which are more stable and for which exciton emission is the main radiative recombination channel. The ASE from such a system cannot be considered as a band-to-band emission and compared with the material gain calculations obtained with the presented approach. Therefore, in comparison with experimental results, we must limit to 3D perovskites where exciton emission at room temperature is not dominant, and \blueB{some of} the selected perovskites belong to such a group. \blueB{This also means that we cannot compare our calculations with ASE measurements for 3D perovskites when the nature of the ASE is excitonic.}

When it comes to 3D perovskites, a primary challenge lies in the material stability, resulting in a limited number of literature reports on gain measurements in this type of material. \blueB{In Ref.~\cite{Alvarado2021}}, systematic gain measurements have been conducted for MAPbI$_{3}$. While a direct comparison between experimentally obtained gain values and those calculated in this study might be challenging due to the impulse excitation of the sample, a qualitative comparison indicates a strong agreement.
The experimental data and the \blueB{results of} theoretical calculations are depicted in Figure~\ref{figGP}. Previous calculations were conducted assuming a broadening of optical transitions with a half-width $\gamma$ equal to \blueB{$0.01$} eV. Figure \ref{figGP} (a) illustrates the maximum gain as a function of the input carrier concentration for various values of $\gamma$, specifically: 0.01, 0.02, 0.03, and 0.04~eV. One can see that as the $\gamma$ parameter increases, the gain values decrease. However, these differences are not very significant. These calculations aim to fit the experimental spectra with the theoretical curves.
The provided values for lasing pumping thresholds $I=50$, $67$, $100$~$\mu$Jcm$^{-2}$ are used to obtain 3D \blueB{carrier} concentration values. \blueB{We estimate the latter with} the relationship 
\blueB{$$n=(1-r)\frac{\beta I}{E_{\mathrm{pump}} \, d},$$}
where $E_{\text{pump}}$ corresponds to the energy of the laser pump wavelength $\lambda=532$ nm \blueB{($E=\frac{\hbar c}{\lambda}$)}, $d$ represents the thin film thickness, \blueB{and $r$ is the part of the light reflected from the perovskite ($r$ for 532 nm is from the range of 0.2-0.4~\cite{Wenwu2017,Park2019,Zhang2019,Samsonova2023}).} 
In Figure \ref{figGP} (a), the large black-filled squares represent the concentrations calculated for \blueB{$r=0.3$ and $\beta=0.35$. $\beta=1$ corresponds to a situation in which each absorbed photon participates in the optical gain. In reality, this is not the case, because some of the absorbed photons form carriers that recombine non-radiatively, and some of the carriers form excitons. At high carrier concentrations, Auger processes begin to play an important role. Therefore, $\beta$ may vary with the excitation density and may be much smaller than $1$ for higher excitation densities. As we take $\beta$ as a constant, it can be treated as a scaling factor when comparing the experimental results with our calculations.} Noticeably, with this scaling factor, one can see a deviation for the highest concentration corresponding to $I = 100\mu$Jcm$^{-2}$. This deviation could be attributed to Auger processes and material degradation.
Figure\ref{figGP} (b) displays gain calculations (solid lines) and experimental data points: red ($I=100\mu$ Jcm$^{-2}$), green ($I=67\mu$ Jcm$^{-2}$), and blue ($I=50\mu $Jcm$^{-2}$). Calculations were performed for $\gamma= 10$~meV. The results of simulations aimed to reproduce the experimental data are plotted in colors corresponding to the relevant data points (red, green, and blue). These dependencies correspond to the input carrier concentrations: $n=1.36\times10^{18}$, $1.42\times10^{18}$, and $1.57\times10^{18}$ cm$^{-3}$.

 \par
 We can therefore conclude that our theoretical predictions are reliable, excluding CsPbI$_{3}$ and CsPbBr$_{3}$, which are in the orthorhombic phase at room temperature, and the cubic phase for these crystals can be obtained at much higher temperatures. In general, \blueB{calculating} the orthorhombic phase for these two materials is possible, but is beyond the scope of this article. 
It should also be mentioned that current laser structures based on III-V materials contain quantum wells because a significant enhancement of material gain takes place as a result of lowering the dimensionality of the system. A similar effect can be expected for quantum wells produced on the basis of 3D perovskites, which are fundamentally different from 2D perovskites. However, calculations of this type of systems are also beyond the scope of this article, \blue{and there is not much work on quantum wells fabricated from 3D perovkites \cite{Parrott2019}}.
\begin{figure}
		\begin{center}
			\includegraphics[width=0.5\textwidth]{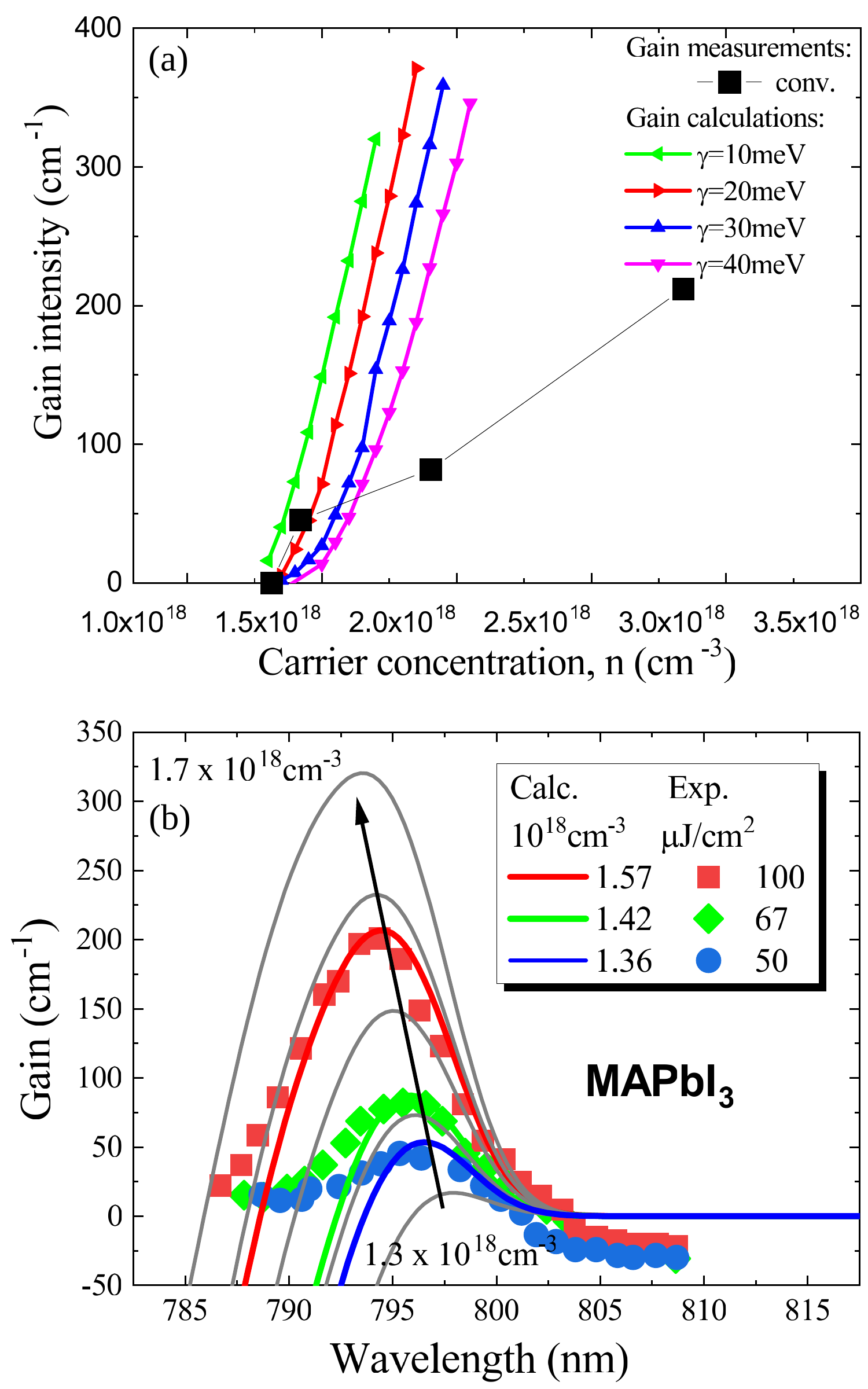}
		\end{center}
		\blue{\caption{\label{figGP} Comparison between MAPbI$_3$ gain calculation and experimental results \cite{Alvarado2021}.
(a) Gain intensity as a function of carrier concentration. The maximum gain value for different $\gamma$ values: 10 meV (green line), 20 meV (red line), 30 meV (blue line), and 40 meV (pink line). The black squares correspond to carrier concentrations calculated from $I=50\mu$Jcm$^{-2}$, $67\mu$Jcm$^{-2}$, and $100\mu$Jcm$^{-2}$ \blueB{using the $\beta$ factor}.
(b) Gain calculation for carrier concentrations ranging from $1.3\times10^{18}$cm$^{-3}$ to $1.7\times10^{18}$cm$^{-3}$. The points correspond to experimental results: red ($I=100\mu$Jcm$^{-2}$), green ($I=67\mu$Jcm$^{-2}$), and blue ($I=50\mu$Jcm$^{-2}$). } }
	\end{figure}

	\section{Conclusions}
	\label{sec:concl}
	In conclusion, we derived an invariant expansion form of the C$_{4\mathrm{v}}$ eight-band $\kp$ Hamiltonian \blueB{for perovskites. We} performed modeling for inorganic and organic halide perovskites of cubic and pseudo-cubic phases. With the obtained results, we calculated the material gain. It has been shown that from the point of view of the electronic band structure, the inorganic (CsPbCl$_3$) and organic (MAPbCl$_3$, MAPbBr$_3$, and MAPbI$_3$) halide perovskites are a very promising gain medium for lasers. A positive material gain for these crystals appears at lower carrier densities than for the reference III-V semiconductors (InP and GaAs) and it is significantly greater.

	\acknowledgments
         We are grateful to Herbert Mączko for sharing his code for material gain calculation for InP and GaAs and to Paweł Scharoch for fruitfull discussions.
	
	\appendix
	\section{Explicit form of the Hamiltonian in the JM basis}
 	\label{sec:app_ham}

	In this appendix, we present an explicit form of the Hamiltonian. The matrix is given in the $\ket{j m}$ basis of the total angular momentum
	$\Big \{  \ket{\frac{1}{2};\frac{1}{2}}_v$,
	$\ket{\frac{1}{2};-\frac{1}{2}}_v$,
	$\ket{\frac{3}{2};\frac{3}{2}}_c$,
	$\ket{\frac{3}{2};\frac{1}{2}}_c$,
	$\ket{\frac{3}{2};-\frac{1}{2}}_c$,
	$\ket{\frac{3}{2};-\frac{3}{2}}_c$,
	$\ket{\frac{1}{2};\frac{1}{2}}_c$,
	$\ket{\frac{1}{2};-\frac{1}{2}}_c \Big \}$. We took the basis definition following Ref.~\cite{Winkler2003}, which results in a consistent form of the invariant matrices, despite the fact that valence and conduction bands are inverted. The basis states in the product form are given by
 \begingroup
	\allowdisplaybreaks
	\begin{align*}
		\ket{\frac{1}{2};\frac{1}{2}}_v &=  \ket{S} \otimes \ket{\uparrow}, \\
            \ket{\frac{1}{2};-\frac{1}{2}}_v &=  \ket{S} \otimes \ket{\downarrow}, \\
		  \ket{\frac{3}{2};\frac{3}{2}}_c &=  -\frac{1}{\sqrt{2}} \ket{X+iY} \otimes \ket{\uparrow}, \\
    	\ket{\frac{3}{2};\frac{1}{2}}_c &=  \frac{2}{\sqrt{6}} \ket{Z} \otimes \ket{\uparrow} - \frac{1}{\sqrt{6}} \ket{X+iY} \otimes \ket{\downarrow}, \\
		\ket{\frac{3}{2};-\frac{1}{2}}_c &=  \frac{1}{\sqrt{6}} \ket{X-iY} \otimes \ket{\uparrow} + \frac{2}{\sqrt{6}} \ket{Z} \otimes \ket{\downarrow}, \\
		\ket{\frac{3}{2};-\frac{3}{2}}_c &=  \frac{1}{\sqrt{2}} \ket{X-iY} \otimes \ket{\downarrow}, \\
		\ket{\frac{1}{2};\frac{1}{2}}_c &=  -\frac{1}{\sqrt{3}} \ket{Z} \otimes \ket{\uparrow} - \frac{1}{\sqrt{3}} \ket{X+iY} \otimes \ket{\downarrow}, \\
		\ket{\frac{1}{2};-\frac{1}{2}}_c &=  -\frac{1}{\sqrt{3}} \ket{X-iY} \otimes \ket{\uparrow} + \frac{1}{\sqrt{3}} \ket{Z} \otimes \ket{\downarrow},\\
	\end{align*}
	\endgroup
         where $S$, $X$, $Y$, $Z$ labels refer to the transformational properties under the symmetry operations.
         According to the common convention, the states that are odd under the inversion operation are taken as purely imaginary, and the even ones are real~\cite{Winkler2003}.
	
	\begin{widetext}
		\begin{equation}
			\label{eq:Hexplicit}
			H=
			\left(
			\begin{array}{*{8}{c}}
				E_{\mathrm{v}} & 0 & -\sqrt{3} V & \sqrt{2} U & V^* & 0 & -U & -\sqrt{2} V^* \\
				0 & E_{\mathrm{v}} & 0 & -V & \sqrt{2} U & \sqrt{3} V^* & -\sqrt{2} V & U \\
				-\sqrt{3} V^* & 0 & E_{\mathrm{hc}} & -S^* & R^* & 0 & \frac{1}{\sqrt{2}} S^* & -\sqrt{2} R^* \\
				\sqrt{2} U^* & -V^* & -S & E_{\mathrm{lc}} & 0 & R^* & -D & -\sqrt{\frac{3}{2}} S^* \\
				V & \sqrt{2} U^* & R & 0 & E_{\mathrm{lc}} & S^* & -\sqrt{\frac{3}{2}} S & D \\
				0 & \sqrt{3} V & 0 & R & S & E_{\mathrm{hc}} & \sqrt{2} R & \frac{1}{\sqrt{2}} S \\
				-U^* & -\sqrt{2} V^* & \frac{1}{\sqrt{2}} S & -D & -\sqrt{\frac{3}{2}} S^* & \sqrt{2} R^* & E_{\mathrm{sc}} & 0 \\
		     	-\sqrt{2} V & U^* & -\sqrt{2} R & -\sqrt{\frac{3}{2}} S & D & \frac{1}{\sqrt{2}} S^* & 0 & E_{\mathrm{sc}} 
			\end{array}
			\right),
		\end{equation}
	\end{widetext}
where
\begingroup
\allowdisplaybreaks
\begin{align*}
	E_{\mathrm{v}} = &  \frac{\hbar^2}{2m_0} k^2, \\
	E_{\mathrm{hc}} = &  E_\mathrm{g} + \Delta_\mathrm{c} + \frac{\delta}{3} +\frac{\hbar^2}{2m_0} \qty[ \gamma'_1 k^2 + \gamma'_2 \qty(k^2_x + k^2_y - 2 k^2_z) ], \\
	E_{\mathrm{lc}} = &  E_\mathrm{g} + \Delta_\mathrm{c} - \frac{\delta}{3} +\frac{\hbar^2}{2m_0} \qty[ \gamma'_1 k^2 - \gamma'_2 \qty(k^2_x + k^2_y - 2 k^2_z) ], \\
	E_{\mathrm{sc}} = &  E_\mathrm{g} + \frac{\hbar^2}{2m_0} \gamma'_1 k^2,  \\
	V = &  \frac{1}{\sqrt{6}} P_{\parallel} \qty(k_x + i k_y),\\
	U = &  \frac{1}{\sqrt{3}} \qty( P_{z} k_z + i \zeta ),\\
	S = & \frac{\hbar^2}{2m_0} 2 \gamma'_3 \sqrt{3} \qty(  k_x + i k_y ) k_z,\\
	R = & -\frac{\hbar^2}{2m_0} \sqrt{3} \qty[ \gamma'_2 \qty(k^2_x - k^2_y) + 2 i \gamma'_3 k_x k_y ],\\
	D = & -\frac{\hbar^2}{2m_0} \sqrt{2} \gamma'_2 \qty(k^2_x + k^2_y - 2 k^2_z).
\end{align*}
\blue{One should note, that the band order in this matrix form is different than in Eq.~\ref{eq:Hblock}.}
\section{Hole effective mass}
\label{sec:app_mass}
\begin{table}[h]
\centering
\caption{Effective masses calculated using the $\kp$ parameters, and fitted directly to the DFT band structures}
\begin{ruledtabular}
\begin{tabular}{lLLL}
\label{tab:mass}
           & \multicolumn{1}{c}{$\kp$} & \multicolumn{2}{c}{Fitted effective masses} \\
           & & \multicolumn{1}{c}{$\langle111\rangle$} & \multicolumn{1}{c}{$\langle010\rangle$} \\[2pt]
           \hline\\[-6pt]
CsPbCl$_3$ &  0.189 & 0.188 & 0.188 \\
CsPbBr$_3$ &  0.158 & 0.151 & 0.151 \\
CsPbI$_3$ &  0.125 & 0.118 & 0.118  \\
MAPbCl$_3$ &  0.210 & 0.208 & 0.228 \\
MAPbBr$_3$ &  0.178 & 0.168 & 0.187 \\
MAPbI$_3$ &  0.141 & 0.133 & 0.151 
\end{tabular}
\end{ruledtabular}
\end{table}
\blue{ In this section, we present a comparison for the valence band effective masses $m_\mathrm{v}$ calculated using the perturbative formula of Eq.~\ref{f2} and the values obtained by fitting to the DFT band structures near $R$-point. The results are given in Table~\ref{tab:mass}. In the case of cubic materials CsPbX$_3$ the DFT data are fitted using a single parabola. However, for the pseudocubic MAPbX$_3$ the valence band splits due to the Rashba coupling~\cite{Even2015} and the fitting is performed using shifted parabolas \blueB{(see Fig.~\ref{fig:splittings})}.
The considered effective masses are in $\langle111\rangle$ (R -- $\Gamma$) and $\langle010\rangle$ (R -- M) directions. In all the cases, the discrepancies between the methods are not very large. Also, the anisotropy is not very pronounced for the considered directions.}
	

\section{The Rashba splitting}
\label{sec:app_rashba}
\blueB{
In this section, we present the derivation of the Rashba parameters within the standard L\"owdin perturbation theory~\cite{Lowdin1950,Winkler2003,LewYanVoon2009}. The two-band model for the valence band can be obtained from the eight-band $\kp$ by decoupling the $\Gamma^-_{8c}$ and $\Gamma^-_{6c}$ band blocks, perturbatively. Neglecting the (small) splitting related to $\delta$ parameter, one can start with the expression
\begin{align*}
    H_\mathrm{6v6v} & =  \frac{\hbar^2}{2 m_0} k^2 +  \frac{H_\mathrm{6v8c} H_\mathrm{8c6v}}{E_{6\mathrm{v}} - E_{8\mathrm{c}}} + 
    \frac{H_\mathrm{6v6c} H_\mathrm{6c6v}}{E_{6\mathrm{v}} - E_{6\mathrm{c}}}.
\end{align*}
Considering the terms linear in $\bm{k}$, one obtains
\begin{align*}
    H^{(\bm{k})}_\mathrm{6v6v} & = \frac{3i \zeta P_\parallel}{E_\mathrm{g} + \Delta_\mathrm{c}} \qty{ \qty(T_x T^\dagger_z - T_z T^\dagger_x) k_x + \qty(T_y T^\dagger_z - T_z T^\dagger_y) k_y} \\ & + \frac{i \zeta P_\parallel}{3 E_\mathrm{g}} \qty( \qty[\sigma_x, \sigma_z] k_x + \qty[\sigma_y, \sigma_z] k_y)
\end{align*}
where the following relations are utilized
        \begin{align*}
            [\sigma_i, \sigma_j] &= 2 i \sum_k \epsilon_{ijk} \sigma_k, \\
            T_i T^\dagger_j &=  \frac{2}{9} \mathbb{I}_2 \delta_{ij} - \frac{i}{9}  \sum_k \epsilon_{ijk} \sigma_k, \\
        \end{align*}	
with $\epsilon_{ijk}$ denoting the Levi-Civita symbol. The further algebra leads to the well-known Rashba Hamiltonian
\begin{align*}
    H^{(\bm{k})}_\mathrm{6v6v} & = \alpha_\mathrm{v} \qty( \sigma_x k_y - \sigma_y k_x)
\end{align*}
with the Rashba parameter $$\alpha_\mathrm{v} = -\frac{2 \zeta P_\parallel \Delta_\mathrm{c} }{3 E_\mathrm{g}(E_\mathrm{g} + \Delta_\mathrm{c})}. $$
Similar calculations that are performed for the lowest conduction band $\Gamma^-_{6c}$ yield
\begin{align*}
    H^{(\bm{k})}_\mathrm{6c6c} & = \alpha_\mathrm{c} \qty( \sigma_x k_y - \sigma_y k_x)
\end{align*}
with $$\alpha_\mathrm{c} = -\frac{2 \zeta P_\parallel}{3 E_\mathrm{g}}. $$
Up to phase convention, the formulas for $\alpha_\mathrm{v/c}$ are consistent with Ref.~\cite{Yu2016} for the limiting case of $\delta \ll \Delta_\mathrm{c}$.
Since the Rashba Hamiltonian can be written in the form $H = \bm{v} \cdot \bm{\sigma}$, its eigenvalues are $\pm \abs{\bm{v}}$, which gives $\pm \alpha_\mathrm{c/v} \sqrt{k^2_x + k^2_y}$. }

	\bibliographystyle{prsty}
	\bibliography{abbr,KG.bib,marta.bib,MW.bib}

\end{document}